\documentclass[a4paper,onecolumn,accepted=2020-09-22]{quantumarticle}
\pdfoutput=1
\usepackage[utf8]{inputenc}
\usepackage[english]{babel}
\usepackage[T1]{fontenc}
\usepackage{amsmath}
\usepackage{hyperref}
\usepackage{tikz}
\usepackage{lipsum}
\usepackage{amsfonts,url}

\newtheorem{theorem}{Theorem}
\newtheorem{definition}{Definition}
\newtheorem{conjecture}{Conjecture}
\newtheorem{lemma}{Lemma}

\begin{document}
\begin{flushright}
YITP-19-120
\end{flushright}

\title{Additive-error fine-grained quantum supremacy}
\author{Tomoyuki Morimae}
\email{tomoyuki.morimae@yukawa.kyoto-u.ac.jp}
\affiliation{Yukawa Institute for Theoretical Physics,
Kyoto University, Kitashirakawa Oiwakecho, Sakyoku, Kyoto
606-8502, Japan}
\affiliation{JST, PRESTO, 4-1-8 Honcho, Kawaguchi, Saitama,
332-0012, Japan}
\author{Suguru Tamaki}
\email{tamak@sis.u-hyogo.ac.jp}
\affiliation{School of Social Information Science, 
University of Hyogo, 8-2-1,
Gakuennishi-machi, Nishi-ku, Kobe, Hyogo 651-2197, Japan}

\maketitle
\begin{abstract}
It is known that several sub-universal
quantum computing models, such as the IQP model,
the Boson sampling model, the one-clean qubit model, and the
random circuit model,
cannot be classically simulated in polynomial time 
under certain conjectures in classical complexity 
theory.
Recently, these results have been improved to ``fine-grained"
versions where even exponential-time classical simulations are
excluded assuming certain classical fine-grained complexity conjectures.
All these fine-grained results are, however,
about the hardness of strong simulations or multiplicative-error
sampling. It was open whether any fine-grained quantum supremacy
result can be shown for a more realistic setup, namely,
additive-error sampling. 
In this paper, we show the additive-error
fine-grained quantum supremacy (under certain
complexity assumptions).
As examples, we consider the IQP model, a mixture of
the IQP model and log-depth Boolean circuits,
and Clifford+$T$ circuits.
Similar results should hold for other sub-universal models.
\end{abstract}


\section{Introduction}
Classically sampling output probability distributions of
sub-universal quantum computing models is known to be
impossible under certain classical complexity conjectures.
The depth-four model~\cite{TD},
the Boson Sampling model~\cite{BS}, 
the IQP model~\cite{BJS,BMS},
the one-clean qubit 
model~\cite{KL,MFF,M,Kobayashi,KobayashiICALP},
the HC1Q model~\cite{HC1Q},
and the random circuit model~\cite{random,random2,random3}
are known examples.
These results prohibit only polynomial-time classical sampling,
but recently,
impossibilities of some
exponential-time classical simulations
have been shown based on classical fine-grained complexity
conjectures~\cite{Dalzell,DalzellPhD,Huang,Huang2,MorimaeTamaki,Hayakawa}.

These ``fine-grained" quantum supremacy results are, however, only for
exact computations (i.e., strong simulations) or multiplicative-error sampling
of output probability distributions. Here, we say that a
quantum probability distribution $\{p_z\}_z$ is classically
sampled in time $T$ within a multiplicative error $\epsilon$ if
there exists a classical $T$-time probabilistic algorithm that outputs
$z$ with probability $q_z$
such that $|p_z-q_z|\le \epsilon p_z$ for all $z$.
It was open whether fine-grained quantum supremacy
can be shown for additive-error sampling. Here,
we say that a
quantum probability distribution $\{p_z\}_z$ is classically
sampled in time $T$ within an additive error $\epsilon$ if
there exists a classical $T$-time probabilistic algorithm that outputs
$z$ with probability $q_z$ such that $\sum_z|p_z-q_z|\le \epsilon$.
Additive-error sampling is more realistic
for medium-size noisy quantum computers, and therefore theoretically showing 
quantum supremacy for additive-error sampling is important for
the near-term experimental demonstrations of quantum supremacy.

In this paper, we show additive-error
fine-grained quantum supremacy based on certain
classical
fine-grained complexity conjectures.
As examples, we consider the IQP model (Sec.~\ref{sec:IQP}), a mixture of
the IQP model and log-depth Boolean circuits (Sec.~\ref{sec:log}),
and Clifford+$T$ circuits (Sec.~\ref{sec:T}).
Similar results should hold for other sub-universal models.

The second result (IQP plus log-depth Boolean circuit) needs more complicated
quantum circuit than the first one, but the conjecture seems to be
more reliable. The first and second results are about the scaling for
the number of qubits, while the third result is about the scaling
for the number of $T$ gates.

The standard proof technique of 
additive-error quantum supremacy~\cite{BS,BMS},
namely, the combination of Markov's inequality,
Stockmeyer's theorem, and the anti-concentration lemma,
cannot be directly used for fine-grained quantum supremacy,
because Stockmeyer's theorem is a result for
polynomial-time probabilistic computing. 
(Markov's inequality and the anti-concentration lemma can be used because
they are independent of the time complexity.)
In order to show fine-grained additive-error quantum supremacy,
we derive a ``fine-grained version" of Stockmeyer's theorem.
Our results, for the first time, demonstrate that the standard proof technique
of additive-error quantum supremacy can be extended
to exponential-time hardness.

Because there is a gap between the upper-bound and the lower-bound, 
the results have a potential to be improved or sharpened. 
The upper-bound can be improved by faster simulations, and the lower-bound can 
be improved by improving the reduction or by introducing other conjectures. 
Furthermore, possible ways are extensions of the 
MA algorithm to refute SETH~\cite{WilliamsCCC}, 
and the fine-grained reductions from approximate counting to decision~\cite{Dell}.

Note: After uploading this paper on arXiv, authors of Ref.~\cite{Dalzell}
told us that they also show independently 
additive-error fine-grained quantum supremacy results.
(Their additive-error results are added in their latest version.)
They consider an exponential-time version of the ${\rm SBP}\neq{\rm SBQP}$
conjecture. (If ${\rm SBQP}={\rm SBP}$, the polynomial-time
hierarchy collapses to the second level.)
At this moment, we do not know how their conjecture
and ours are related.

\section{IQP}
\label{sec:IQP}
In this section, we show additive-error fine-grained quantum supremacy
of the IQP model. 
The IQP model is defined as follows.

\begin{definition}
An $N$-qubit IQP model is the following quantum computing model:
\begin{itemize}
\item[1.]
The initial state is $|0^N\rangle$.
(Here, $|0^N\rangle=|0\rangle^{\otimes N}$.)
\item[2.]
$H^{\otimes N}$ is applied,
where $H$ is the Hadamard gate.
\item[3.]
$Z$-diagonal gates (such as $e^{i\theta Z}$, $Z$, $CZ$, and $CCZ$) are applied.
(In this paper, we consider only $Z$, $CZ$, and $CCZ$.)
\item[4.]
$H^{\otimes N}$ is applied.
\item[5.]
All qubits are measured in the computational basis.
\end{itemize}
\end{definition}

Let us
consider an $n$-variable
degree-3 polynomial, $f:\{0,1\}^n\to\{0,1\}$, 
over ${\mathbb F}_2$
defined by
\begin{eqnarray*}
f(x_1,...,x_n)\equiv
\sum_{i=1}^n \alpha_i x_i
+\sum_{i>j}\beta_{i,j} x_ix_j
+\sum_{i>j>k}\gamma_{i,j,k} x_ix_jx_k
\end{eqnarray*}
for any $x\equiv(x_1,x_2,...,x_n)\in\{0,1\}^n$,
where 
$\alpha_i,\beta_{i,j},\gamma_{i,j,k}\in\{0,1\}$.
If we say that we randomly choose $f$, it means that
we 
randomly choose
each $\alpha_i,\beta_{i,j},\gamma_{i,j,k}$ uniformly and independently.

The conjecture on which  additive-error fine-grained quantum supremacy
of the IQP model is based is stated as follows.

\begin{conjecture}
\label{conjecture:IQP}
Let $f$ be an $n$-variable degree-3 polynomial
over ${\mathbb F}_2$.
Let us define
\begin{eqnarray*}
gap(f)\equiv\sum_{x\in\{0,1\}^n}(-1)^{f(x)}.
\end{eqnarray*}
There exist positive constants $a$ and $n_0$ such that
for every $n>n_0$ the following holds.
Computing $[gap(f)]^2$
within a multiplicative error $u$ for
at least $v$ fraction of $f$ cannot be done
with a classical probabilistic $O^*(2^{a n})$-time
algorithm that
makes 
queries of length $O(2^{an})$ to
an ${\rm NTIME}[n^2]$ oracle 
with a success probability at least $w$.
Here, $u,v,w$ are certain constants.
($O^*$ means that the polynomial factor is ignored.)
\end{conjecture}

Here, the oracle query is the standard one: there is a separate
oracle tape and answers can be returned from the oracle instantaneously.
Note that the parameters $u,v,w$ 
can be adjustable to some extent. (See the proof.)
We do not know whether this conjecture is true or false, but at least
at this moment we do not know how to refute it.
(For more discussions, see Sec.~\ref{sec:conjectures}).

Based on Conjecture~\ref{conjecture:IQP}, we show the following result.

\begin{theorem}
\label{theorem:IQP}
If Conjecture~\ref{conjecture:IQP} is true, then
there exists an $N$-qubit IQP circuit whose output probability distribution
cannot be classically sampled in $O(2^{aN})$-time
within a certain constant additive error $\epsilon$.
\end{theorem}

For simplicity, we consider
degree-3 polynomials, but
it is clear from the following proof
that a similar result holds for degree-$k$ polynomials
for any constant $k\ge3$.
(The anti-concentration lemma holds for any degree-$k$ polynomial
with $k\ge2$, but the degree-2 case is classically simulatable because it
is a Clifford circuit, so
$k\ge3$ is necessary.)
If we consider Conjecture~\ref{conjecture:IQP} for
degree-$k$ polynomials, it becomes more stable for larger 
$k$~\cite{Williams}.

{\it Proof of Theorem~\ref{theorem:IQP}}.
Given an $n$-variable degree-3 polynomial $f$,
we can construct an $n$-qubit IQP circuit such that
the probability $p_z(f)$ of outputting $z\in\{0,1\}^n$ satisfies
\begin{eqnarray*}
p_z(f)=\frac{(gap(f_z))^2}{2^{2n}},
\end{eqnarray*}
where
\begin{eqnarray*}
f_z(x_1,...,x_n)\equiv f(x_1,...,x_n)+\sum_{i=1}^n z_ix_i.
\end{eqnarray*}

Assume that there exists a $T$-time
classical probabilistic algorithm that
outputs $z\in\{0,1\}^n$ with probability $q_z(f)$ such that
\begin{eqnarray*}
\sum_{z\in\{0,1\}^n}|p_z(f)-q_z(f)|\le\epsilon
\end{eqnarray*}
for a certain $\epsilon$ and any $f$.
From Markov's inequality,
\begin{eqnarray*}
{\rm Pr}_z
\Big[|p_z(f)-q_z(f)|\ge\frac{\epsilon}{2^n\delta}\Big]
\le \delta
\end{eqnarray*}
for any $f$ and $\delta>0$.
According to the fine-grained Stockmeyer's theorem 
(see Appendix),
a classical $O^*(T)$-time probabilistic algorithm 
that makes queries of $O(T)$ length to
the ${\rm NTIME}[n^2]$ oracle 
can compute $\tilde{q}_z(f)$ such that
\begin{eqnarray*}
|q_z(f)-\tilde{q}_z(f)|\le 
\xi q_z(f),
\end{eqnarray*}
where
\begin{eqnarray*}
\xi\equiv
\frac{2^{\frac{1}{\alpha}}-2^{-\frac{1}{\alpha}}}{2},
\end{eqnarray*}
for any $f$, integer
$\alpha\ge1$, and $z\in\{0,1\}^n$, with a success probability
at least $w$.
Due to the anti-concentration lemma~\cite{BMS}, 
\begin{eqnarray*}
{\rm Pr}_{z,f}\Big[p_z(f)\ge\frac{\tau}{2^n}\Big]
\ge \frac{(1-\tau)^2}{3}
\end{eqnarray*}
for any $0<\tau<1$.

Therefore we have
\begin{eqnarray*}
|p_z(f)-\tilde{q}_z(f)|&\le&|p_z(f)-q_z(f)|
+|q_z(f)-\tilde{q}_z(f)|\\
&\le&|p_z(f)-q_z(f)|+\xi q_z(f)
~~\mbox{(with a success probability at least $w$ for each $f$
and $z$)}\\
&\le&|p_z(f)-q_z(f)|+\xi(p_z(f)+|p_z(f)-q_z(f)|)\\
&=&\xi p_z(f)+|p_z(f)-q_z(f)|(1+\xi)\\
&\le&\xi p_z(f)
+\frac{\epsilon}{2^n\delta}(1+\xi)
~~\mbox{(for at least $1-\delta$ fraction of $z$)}\\
&\le&\xi p_z(f)+\sigma p_z(f)(1+\xi)
~~\mbox{(for at least 
$\frac{(1-\frac{\epsilon}{\sigma\delta})^2}{3}$ 
fraction of $(z,f)$)}\\
&=&p_z(f)\Big(\sigma+(1+\sigma)\xi\Big)\\
&=&p_z(f)u~~
(\mbox{We take $u\equiv \sigma+(1+\sigma)\xi$}).
\end{eqnarray*}
If we take $\epsilon$ and $\delta$ such that
$-\delta+\frac{1}{3}(1-\frac{\epsilon}{\sigma\delta})^2=v$,
the above inequality is correct for at least
$v$
fraction of $(z,f)$.
Hence, we obtain
\begin{eqnarray*}
|(gap(f_z))^2-2^{2n}\tilde{q}_z(f)|\le u(gap(f_z))^2
\end{eqnarray*}
for at least $v$ fraction of $(z,f)$.
It means
$(gap(f))^2$ is computable within the multiplicative error
$u$
for at least $v$ fraction of $f$,
which contradict Conjecture~\ref{conjecture:IQP}.
\fbox

Note that $w$ can be arbitrary
close to 1, but $u$ is lower-bounded as
$u\ge\frac{\epsilon}{1+\sqrt{3}}$,
and $v$ is upper-bounded as
$v \le1-\frac{\epsilon}{u(1+\sqrt{3})}$.

\section{IQP plus log-depth Boolean circuit}
\label{sec:log}
In this section,
we show additive-error fine-grained quantum supremacy
for the IQP plus log-depth Boolean circuit model.

Let us consider the following conjecture.
\begin{conjecture}
\label{conjecture:log}
Let 
$f:\{0,1\}^n\to\{0,1\}$ be an $n$-variable degree-2 polynomial over
${\mathbb F}_2$,
and $g:\{0,1\}^n\to\{0,1\}$ be an $n$-variable log-depth Boolean circuit. 
Define
\begin{eqnarray*}
gap(f+g)\equiv\sum_{x\in\{0,1\}^n}(-1)^{f(x)+g(x)}.
\end{eqnarray*}
There exist $g$, and positive constants $a$ and $n_0$ such that for
every $n>n_0$ the following holds.
Computing $[gap(f+g)]^2$
within a multiplicative error
$u$ for at least $v$ fraction of
$f$ cannot be done with a classical probabilistic 
$O^*(2^{an})$-time
algorithm that makes queries of length
$O(2^{an})$ 
to an ${\rm NTIME}[n^2]$ oracle
with a success probability 
at least $w$.
\end{conjecture}

Conjecture~\ref{conjecture:log} is ``more stable" than
Conjecture~\ref{conjecture:IQP}, because log-depth Boolean circuit
is more general than constant-degree polynomials.
For constant-degree polynomials, there is a non-trivial exponential
time algorithm to count the number of solutions~\cite{Tamaki}, but
we do not know how to apply it to log-depth Boolean circuits.
Furthermore, note that in Conjecture~\ref{conjecture:log},
the average case is considered only for $f$, and $g$ can be taken
as the worst case one.

Based on Conjecture~\ref{conjecture:log}, we show the following result.

\begin{theorem}
\label{theorem:log}
If Conjecture~\ref{conjecture:log} is true,
then there exists an $N$-qubit $poly(N)$-size
quantum circuit (consisting of an IQP circuit and a log-depth Boolean circuit) 
whose output probability distribution
cannot be classically sampled
in $O(2^{aN})$-time within a 
certain constant additive error $\epsilon$.
\end{theorem}

{\it Proof of Theorem~\ref{theorem:log}.}
Given a log-depth Boolean circuit $g:\{0,1\}^n\to\{0,1\}$, we can construct
an $(n+1)$-qubit $poly(n)$-size quantum circuit $U$ such that
\begin{eqnarray*}
U(|x\rangle\otimes|0\rangle)
=e^{ih(x)}|x\rangle\otimes|g(x)\rangle
\end{eqnarray*}
for any $x\in\{0,1\}^n$, where $h$ is a certain function
whose detail is irrelevant here~\cite{Cosentino}.
Let us consider the following circuit.
\begin{itemize}
\item[1.]
The initial state is $|0^n\rangle\otimes|0\rangle$.
\item[2.]
Apply $H^{\otimes n}\otimes I$ to obtain
\begin{eqnarray*}
\frac{1}{\sqrt{2^n}}\sum_{x\in\{0,1\}^n}|x\rangle\otimes|0\rangle.
\end{eqnarray*}
\item[3.]
Apply $U$ to obtain
\begin{eqnarray*}
\frac{1}{\sqrt{2^n}}\sum_{x\in\{0,1\}^n}e^{ih(x)}
|x\rangle\otimes|g(x)\rangle.
\end{eqnarray*}
\item[4.]
Apply $Z$ on the last qubit to obtain
\begin{eqnarray*}
\frac{1}{\sqrt{2^n}}\sum_{x\in\{0,1\}^n}e^{ih(x)}(-1)^{g(x)}
|x\rangle\otimes|g(x)\rangle.
\end{eqnarray*}
\item[5.]
Apply $U^\dagger$ to obtain
\begin{eqnarray*}
\frac{1}{\sqrt{2^n}}\sum_{x\in\{0,1\}^n}
(-1)^{g(x)}
|x\rangle\otimes|0\rangle.
\end{eqnarray*}
\item[6.]
Apply $Z$ and $CZ$ that correspond to $f$ to obtain
\begin{eqnarray*}
\frac{1}{\sqrt{2^n}}\sum_{x\in\{0,1\}^n}
(-1)^{g(x)+f(x)}
|x\rangle\otimes|0\rangle.
\end{eqnarray*}
\item[7.]
Apply $H^{\otimes n}\otimes I$ and measure 
the first $n$ qubits in the computational basis.
\end{itemize}
The probability of obtaining $z\in\{0,1\}^n$ is
\begin{eqnarray*}
p_z(f+g)
&=&\Big|
\frac{1}{2^n}\sum_{x\in\{0,1\}^n}
(-1)^{g(x)+f(x)+\sum_{j=1}^nx_jz_j}
\Big|^2\\
&=&
\frac{(gap(g+f_z))^2}{2^{2n}}.
\end{eqnarray*}

Assume that there exists a $T$-time
classical probabilistic algorithm that
outputs $z\in\{0,1\}^n$ with probability $q_z(f+g)$ such that
\begin{eqnarray*}
\sum_{z\in\{0,1\}^n}|p_z(f+g)-q_z(f+g)|\le\epsilon.
\end{eqnarray*}
From Markov's inequality,
\begin{eqnarray*}
{\rm Pr}_z
\Big[|p_z(f+g)-q_z(f+g)|\ge\frac{\epsilon}{2^n\delta}\Big]
\le \delta
\end{eqnarray*}
for any $f$, $g$, and $\delta>0$.
According to the fine-grained Stockmeyer's theorem,
a classical $O^*(T)$-time probabilistic algorithm that makes
queries of length $O(T)$ to
the
${\rm NTIME}[n^2]$ oracle
can compute $\tilde{q}_z(f+g)$ such that
\begin{eqnarray*}
|q_z(f+g)-\tilde{q}_z(f+g)|\le \xi q_z(f+g),
\end{eqnarray*}
where
\begin{eqnarray*}
\xi\equiv\frac{2^{\frac{1}{\alpha}}-2^{-\frac{1}{\alpha}}}{2},
\end{eqnarray*}
for any $f$, $g$, integer $\alpha\ge1$, and $z\in\{0,1\}^n$, with a success probability
at least $w$.
Due to the anti-concentration lemma~\cite{BMS} 
\begin{eqnarray*}
{\rm Pr}_{z,f}\Big[p_z(f+g)\ge\frac{\tau}{2^n}\Big]
\ge \frac{(1-\tau)^2}{3}
\end{eqnarray*}
for any $0<\tau<1$.

Then we have
\begin{eqnarray*}
|p_z(f+g)-\tilde{q}_z(f+g)|&\le&|p_z(f+g)-q_z(f+g)|
+|q_z(f+g)-\tilde{q}_z(f+g)|\\
&\le&|p_z(f+g)-q_z(f+g)|+\xi q_z(f+g)\\
&\le&|p_z(f+g)-q_z(f+g)|+\xi(p_z(f+g)+|p_z(f+g)-q_z(f+g)|)\\
&=&\xi p_z(f+g)+|p_z(f+g)-q_z(f+g)|(1+\xi)\\
&\le&\xi p_z(f+g)
+\frac{\epsilon}{2^n\delta}(1+\xi)
~~\mbox{(for at least $1-\delta$ fraction of $z$)}\\
&\le&\xi p_z(f+g)+\sigma p_z(f+g)(1+\xi)
~~\mbox{(for at least $\frac{(1-\frac{\epsilon}{\sigma\delta})^2}{3}$ 
fraction of $(z,f)$)}\\
&=&p_z(f+g)\Big(\sigma+(1+\sigma)\xi\Big)\\
&=&p_z(f+g)u
~~\mbox{(We take $u\equiv\sigma+(1+\sigma)\xi$)}.
\end{eqnarray*}
If we take $\epsilon$ and $\delta$ 
such that 
$-\delta+\frac{1}{3}(1-\frac{\epsilon}{\sigma\delta})^2=v$,
the above inequality is correct for at least $v$ fraction
of $(z,f)$, which contradict Conjecture~\ref{conjecture:log}.
\fbox

\section{Clifford plus $T$}
\label{sec:T}
In this section, we finally show additive-error fine-grained
quantum supremacy for Clifford+$T$ circuits.
Let us consider the following conjecture.
\begin{conjecture}
\label{conjecture:T}
Let 
$g:\{0,1\}^n\to\{0,1\}$ be a 3-CNF with $m$ clauses,
and $f:\{0,1\}^n\to\{0,1\}$ be an $n$-variable degree-2 polynomial
over ${\mathbb F}_2$.
Define
\begin{eqnarray*}
gap(f+g)\equiv\sum_{x\in\{0,1\}^n}(-1)^{f(x)+g(x)}.
\end{eqnarray*}
There exist $g$, and positive constants $a$ and $n_0$ such that for
every $n>n_0$ the following holds.
Computing $[gap(f+g)]^2$
within a multiplicative error
$u$ for at least $v$ fraction of
$f$ cannot be done with a classical probabilistic 
$O^*(2^{am})$-time
algorithm that makes queries 
of length $O(2^{am})$ to
an ${\rm NTIME}[n^2]$ oracle
with a success probability 
at least $w$.
\end{conjecture}

Based on Conjecture~\ref{conjecture:T}, we show the following result.

\begin{theorem}
\label{theorem:T}
If Conjecture~\ref{conjecture:T} is true,
then there exists a
quantum circuit over Clifford gates and
$t$ $T$ gates whose output probability distribution
cannot be classically sampled
in $O(2^{\frac{a(t+14)}{42}})$-time within a 
certain constant additive error
$\epsilon$.
\end{theorem}

{\it Proof of Theorem~\ref{theorem:T}.}
Given a 3-CNF $g:\{0,1\}^n\to\{0,1\}$, we can construct
a quantum circuit $U$ such that
\begin{eqnarray*}
U(|x\rangle\otimes|0^\xi\rangle)
=|g(x)\rangle\otimes|junk(x)\rangle
\end{eqnarray*}
for any $x\in\{0,1\}^n$, where $\xi\equiv 3m-1$,
and
$junk(x)\in\{0,1\}^{n+\xi-1}$
is a certain bit string whose detail is irrelevant here.
Note that $U$
consists of Clifford and $7(3m-1)$ number of $T$ gates.
(The 3-CNF $g$ contains $2m$ OR gates and $m-1$ AND gates.
Each AND and OR gate can be simulated with a single Toffoli gate
by using a single ancilla qubit.
A single Toffoli gate can be simulated with Clifford and 7 $T$ gates.)
Let us consider the following circuit.
\begin{itemize}
\item[1.]
The initial state is $|0^n\rangle\otimes|0^\xi\rangle$.
\item[2.]
Apply $H^{\otimes n}\otimes I^{\otimes \xi}$ to obtain
\begin{eqnarray*}
\frac{1}{\sqrt{2^n}}\sum_{x\in\{0,1\}^n}|x\rangle\otimes|0^\xi\rangle.
\end{eqnarray*}
\item[3.]
Apply $U$ to obtain
\begin{eqnarray*}
\frac{1}{\sqrt{2^n}}\sum_{x\in\{0,1\}^n}
|g(x)\rangle\otimes|junk(x)\rangle.
\end{eqnarray*}
\item[4.]
Apply $Z\otimes I^{\otimes n+\xi-1}$ to obtain
\begin{eqnarray*}
\frac{1}{\sqrt{2^n}}\sum_{x\in\{0,1\}^n}
(-1)^{g(x)}|g(x)\rangle\otimes|junk(x)\rangle.
\end{eqnarray*}
\item[5.]
Apply $U^\dagger$ to obtain
\begin{eqnarray*}
\frac{1}{\sqrt{2^n}}\sum_{x\in\{0,1\}^n}
(-1)^{g(x)}|x\rangle\otimes|0^\xi\rangle.
\end{eqnarray*}
\item[6.]
Apply $Z$ and $CZ$ that correspond to $f$ to obtain
\begin{eqnarray*}
\frac{1}{\sqrt{2^n}}\sum_{x\in\{0,1\}^n}
(-1)^{g(x)+f(x)}|x\rangle\otimes|0^\xi\rangle.
\end{eqnarray*}
\item[7.]
Apply $H^{\otimes n}\otimes I^{\otimes \xi}$ and measure 
all qubits in the first register in the computational basis.
\end{itemize}
This quantum computing uses $t\equiv14(3m-1)$ number of $T$ gates.
The probability of obtaining $z\in\{0,1\}^n$ is
\begin{eqnarray*}
p_z(f+g)
&=&\Big|
\frac{1}{2^n}\sum_{x\in\{0,1\}^n}
(-1)^{f(x)+\sum_{j=1}^nx_jz_j+g(x)}
\Big|^2.
\end{eqnarray*}

Assume that there exists a $T$-time
classical probabilistic algorithm that
outputs $z\in\{0,1\}^n$ with probability $q_z(f+g)$ such that
\begin{eqnarray*}
\sum_{z\in\{0,1\}^n}|p_z(f+g)-q_z(f+g)|\le\epsilon.
\end{eqnarray*}
From Markov's inequality,
\begin{eqnarray*}
{\rm Pr}_z
\Big[|p_z(f+g)-q_z(f+g)|\ge\frac{\epsilon}{2^n\delta}\Big]
\le \delta
\end{eqnarray*}
for any $f$, $g$, and $\delta>0$.
According to the fine-grained Stockmeyer's theorem,
a classical $O^*(T)$-time probabilistic algorithm 
that makes queries of length $O(T)$ to
the
${\rm NTIME}[n^2]$ oracle
can compute $\tilde{q}_z(f+g)$ such that
\begin{eqnarray*}
|q_z(f+g)-\tilde{q}_z(f+g)|\le \xi q_z(f+g),
\end{eqnarray*}
where
\begin{eqnarray*}
\xi\equiv\frac{2^{\frac{1}{\alpha}}-2^{-\frac{1}{\alpha}}}{2},
\end{eqnarray*}
for any $f$, $g$, integer $\alpha\ge1$, and $z\in\{0,1\}^n$, with a success probability
at least $w$.
Due to the anti-concentration lemma~\cite{BMS} 
\begin{eqnarray*}
{\rm Pr}_{z,f}\Big[p_z(f+g)\ge\frac{\tau}{2^n}\Big]
\ge \frac{(1-\tau)^2}{3}
\end{eqnarray*}
for any $0<\tau<1$.

Then we have
\begin{eqnarray*}
|p_z(f+g)-\tilde{q}_z(f+g)|&\le&|p_z(f+g)-q_z(f+g)|
+|q_z(f+g)-\tilde{q}_z(f+g)|\\
&\le&|p_z(f+g)-q_z(f+g)|+\xi q_z(f+g)\\
&\le&|p_z(f+g)-q_z(f+g)|+\xi(p_z(f+g)+|p_z(f+g)-q_z(f+g)|)\\
&=&\xi p_z(f+g)+|p_z(f+g)-q_z(f+g)|(1+\xi)\\
&\le&\xi p_z(f+g)
+\frac{\epsilon}{2^n\delta}(1+\xi)
~~\mbox{(for at least $1-\delta$ fraction of $z$)}\\
&\le&\xi p_z(f+g)+\sigma p_z(f+g)(1+\xi)
~~\mbox{(for at least $\frac{(1-\frac{\epsilon}{\sigma\delta})^2}{3}$ 
fraction of $(z,f)$)}\\
&=&p_z(f+g)\Big(\sigma+(1+\sigma)\xi\Big)\\
&=&p_z(f+g)u
~~\mbox{(We take $u\equiv\sigma+(1+\sigma)\xi$)}.
\end{eqnarray*}
If we take 
$\epsilon$ and $\delta$ such that
$-\delta+\frac{1}{3}(1-\frac{\epsilon}{\sigma\delta})^2=v$,
the above inequality is correct for at least $v$ fraction
of $(z,f)$, which contradict Conjecture~\ref{conjecture:T}.
\fbox

\section{Discussion}
\subsection{Conjectures}
\label{sec:conjectures}
In this paper, we have shown additive-error
fine-grained quantum supremacy based on several conjectures.
In this subsection, we provide some ``evidence" that support these conjectures.

Our conjectures are related to the exponential-time hypothesis (ETH) 
and the strong exponential-time hypothesis (SETH) that are standard conjectures
in fine-grained complexity theory~\cite{Impagliazzo1,Impagliazzo2}.
ETH and SETH are stronger (or more pessimistic) 
versions of the famous ${\rm NP}\neq{\rm P}$ conjecture that says that
an NP-complete problem cannot be solved in polynomial-time.
More precisely, ETH is stated as follows:

\begin{conjecture}(ETH)
Any (classical) deterministic algorithm that decides
whether $\#f>0$ or $\#f=0$ given (a description of) a 3-CNF
with $n$ variables, $f:\{0,1\}^n\to\{0,1\}$, needs
$2^{\Omega(n)}$ time. Here, $\#f\equiv\sum_{x\in\{0,1\}^n}f(x)$.
\end{conjecture}

SETH is the stronger version of ETH which says as follows:

\begin{conjecture}(SETH)
Let $A$ be any (classical) deterministic $T(n)$-time algorithm such that
the following holds: given (a description of) a CNF, 
$f:\{0,1\}^n\to\{0,1\}$, with at most $cn$ clauses, $A$ accepts
if $\#f>0$ and rejects if $\#f=0$, where $\#f\equiv\sum_{x\in\{0,1\}^n}f(x)$.
Then, for any constant $a>0$, there exists a constant $c>0$ such that
$T(n)>2^{(1-a)n}$ holds for infinitely many $n$.
\end{conjecture}

All conjectures used in this paper
are the average-case hardness of computing GapP functions
within a multiplicative error in classical probabilistic time
with an NTIME oracle, and therefore different from ETH and SETH.
There are, however, three reasons that support these conjectures.

First, our conjectures consider GapP functions, while ETH and SETH
consider $\#$P functions.
A GapP function is a difference of two $\#$P functions.
Furthermore,
to our knowledge, only known way of computing a GapP function is to compute
the number of accepting and rejecting paths (i.e., $\#$P functions).
Therefore, computing GapP functions should not be easier than computing
$\#$P functions.

Second, our conjectures study average cases.
One might think that
solving an average case could be easier than the worst case,
but, at least, SETH has not been refuted even in average cases.
(The best upper-bound is Ref.~\cite{Lincoln}.) 

Third, our conjectures allow the algorithm to use an NTIME oracle.
We point out that at least $\#$ETH, which is the counting version
of ETH, has not been refuted for 
MA (which is in ${\rm ZPP}^{\rm NP}$)
and AM (which is ${\rm coNP}^{\rm NP}$).

It is an important open problem for the research of
(not only fine-grained but also non-fine-grained) quantum supremacy 
to show additive-error quantum supremacy based on
standard conjectures.

\subsection{Other models}
For simplicity, we have considered the three models, but similar results
should hold for other sub-universal
models such as the one-clean qubit model and
the random circuit model. For all sub-universal models, Markov's inequality
and the anti-concentration lemma hold. (For the Boson sampling model,
the anti-concentration is a conjecture.)
Therefore if we assume similar average-case hardness conjectures as we have
introduced in this paper,
we should be able to show additive-error fine-grained quantum supremacy
for other sub-universal models.

\acknowledgements
TM thanks Yoshifumi Nakata for discussion.
We thank authors of Ref.~\cite{Dalzell} for comments on our manuscript,
and sharing their draft.
We also thank anonymous reviewers, especially one reviewer
who told us some errors and an improvement of fine-grained Stockmeyer's
theorem.
TM is supported by MEXT Q-LEAP, JST PRESTO No.JPMJPR176A,
and the Grant-in-Aid for Young Scientists (B) No.JP17K12637 of JSPS. 
ST is supported by JSPS KAKENHI Grant Numbers 16H02782, 18H04090, 
and 18K11164.

\appendix
\section{Appendix}
In this Appendix, we provide a proof of the fine-grained
Stockmeyer's theorem.
The proof is a straightforward generalization of
the one given in Ref.~\cite{Trevisan}.

\subsection{A pairwise independent hash family and the leftover
hash lemma}
To show the fine-grained Stockmeyer's theorem, we need the following
two lemmas.
Their proofs can be found in standard text books of complexity theory,
such as Ref.~\cite{Goldreich}.

\begin{lemma}[A pairwise independent hash family]
\label{construction}
Let $A$ be a random $m\times n$ binary Toeplitz matrix, and
$b$ be a random $m$-dimensional binary vector.
(Here, a Toeplitz matrix is a matrix whose matrix elements
satisfy $a_{i,j}=a_{i+1,j+1}$.)
Then, the family $H\equiv\{h_{A,b}\}_{A,b}$ of 
functions, $h_{A,b}:\{0,1\}^n\to\{0,1\}^m$, with
$h_{A,b}(x)\equiv Ax+b$
satisfies
\begin{eqnarray*}
{\rm Pr}_{A,b}[h_{A,b}(x_1)=y_1\wedge h_{A,b}(x_2)=y_2]=\frac{1}{2^{2m}}
\end{eqnarray*}
for any $x_1\neq x_2\in\{0,1\}^n$ and $y_1,y_2\in\{0,1\}^m$.
\end{lemma}

\begin{lemma}[The leftover hash lemma]
\label{S}
Let $S\subseteq\{0,1\}^n$ be a set of $n$-bit strings.
Then
\begin{eqnarray*}
{\rm Pr}_{A,b}\Big[
\Big||\{x\in S:h_{A,b}(x)=0^m\}|-\frac{|S|}{2^m}\Big|
\ge \epsilon\frac{|S|}{2^m}\Big]
\le\frac{2^m}{\epsilon^2|S|}.
\end{eqnarray*}
\end{lemma}

\subsection{Algorithm $A_k$}
In this subsection, we construct the algorithm $A_k$, which is used
for fine-grained Stockmeyer's theorem.
Let $G$ be an $n$-time deterministic classical algorithm.
Let $S\equiv\{x\in\{0,1\}^n:G(x)=1\}$.
Let $k$ be an integer such that $1\le k\le n$.
We construct a classical
probabilistic $O(rn)$-time algorithm 
$A_k$ 
that gets a description of $G$ as the input, and
that makes
queries of length $O(n)$ to
the ${\rm NTIME}[n^2]$ oracle
such that
\begin{itemize}
\item
If $|S|\ge 2^{k+1}$ then 
${\rm Pr}[A_k~{\rm accepts}]\ge1-e^{-r}$.
\item
If $|S|< 2^k$ then 
${\rm Pr}[A_k~{\rm accepts}]\le e^{-r}$.
\end{itemize}

To construct $A_k$, let us consider the following algorithm.
\begin{itemize}
\item[1.]
If $k\le5$, query the ${\rm NTIME}[n]$
oracle whether $|S|\ge2^{k+1}$ or not.
(The query to the oracle is the description of $G$.
Given the description of $G$, deciding $|S|\ge2^{k+1}$ or not is in
${\rm NTIME}[n]$.)
Accept if the oracle answer is yes.
If the oracle answer is no, reject.
\item[2.]
If $k\ge 6$,
set $m\equiv k-5$.
Randomly choose an $m\times n$ binary Toeplitz matrix $A$,
and an $m$-dimensional binary vector $b$. 
It takes $n+2m-1=O(n)$ time.
Define the function $h_{A,b}:\{0,1\}^n\to\{0,1\}^m$
by $h_{A,b}(x)\equiv Ax+b$.
Query the ${\rm NTIME}[n^2]$ 
oracle whether $|\{x\in S:h_{A,b}(x)=0^m\}|\ge 48$ or not.
(The query to the oracle is the description of $G$, $A$, and $b$.
Given the description of $G$, $A$, and $b$, deciding 
$|\{x\in S:h_{A,b}(x)=0^m\}|\ge 48$ or not
is in ${\rm NTIME}[n^2]$.)
If the oracle answer is yes, accept.
If the oracle answer is no, reject.
\end{itemize}

Assume that $k\le 5$.
Then, if $|S|\ge2^{k+1}$, the probability that the
algorithm accepts is 1.
If $|S|<2^k$, the probability that the algorithm accepts is 0.

Assume that $k\ge 6$.
If $|S|\ge 2^{k+1}$, then $|S|\ge 2^{m+6}$ and therefore
\begin{eqnarray*}
{\rm Pr}[{\rm reject}]
&=&
{\rm Pr}_{A,b}\Big[
|\{x\in S:h_{A,b}(x)=0^m\}|< 48
\Big]\\
&\le&
{\rm Pr}_{A,b}\Big[
|\{x\in S:h_{A,b}(x)=0^m\}|\le \frac{3}{4}\frac{|S|}{2^m}
\Big]\\
&=&
{\rm Pr}_{A,b}\Big[
-|\{x\in S:h_{A,b}(x)=0^m\}|+\frac{|S|}{2^m}\ge \frac{1}{4}\frac{|S|}{2^m}
\Big]\\
&\le&
{\rm Pr}_{A,b}\Big[\Big|
|\{x\in S:h_{A,b}(x)=0^m\}|-\frac{|S|}{2^m}\Big|\ge \frac{1}{4}\frac{|S|}{2^m}
\Big]\\
&\le&\frac{2^{m+4}}{|S|}\le\frac{1}{4}.
\end{eqnarray*}
If $|S|<2^k$, define a superset $S'\supseteq S$ with $|S'|=2^k$.
Then,
\begin{eqnarray*}
{\rm Pr}[{\rm accept}]&=&
{\rm Pr}_{A,b}\Big[
|\{x\in S:h_{A,b}(x)=0^m\}|\ge 48\Big]\\
&\le&
{\rm Pr}_{A,b}\Big[
|\{x\in S':h_{A,b}(x)=0^m\}|\ge 48\Big]\\
&=&
{\rm Pr}_{A,b}\Big[
|\{x\in S':h_{A,b}(x)=0^m\}|\ge \frac{3}{2}\frac{|S'|}{2^m}\Big]\\
&=&
{\rm Pr}_{A,b}\Big[
|\{x\in S':h_{A,b}(x)=0^m\}|\ge \frac{|S'|}{2^m}+\frac{1}{2}\frac{|S'|}{2^m}\Big]\\
&=&
{\rm Pr}_{A,b}\Big[
|\{x\in S':h_{A,b}(x)=0^m\}|-\frac{|S'|}{2^m}\ge\frac{1}{2}\frac{|S'|}{2^m}\Big]\\
&\le&
{\rm Pr}_{A,b}\Big[
\Big|
|\{x\in S':h_{A,b}(x)=0^m\}|-\frac{|S'|}{2^m}\Big|\ge\frac{1}{2}\frac{|S'|}{2^m}\Big]\\
&\le&\frac{2^{m+2}}{|S'|}=\frac{1}{8}.
\end{eqnarray*}

Therefore we have constructed the algorithm such that
\begin{itemize}
\item
If $|S|\ge 2^{k+1}$ then 
${\rm Pr}[{\rm accept}]\ge\frac{3}{4}$.
\item
If $|S|< 2^k$ then 
${\rm Pr}[{\rm accept}]\le\frac{1}{8}$.
\end{itemize}
By repeating this algorithm $O(r)$ times, we get the final result.

\subsection{Fine-grained Stockmeyer's theorem}
\label{sec:stockmeyer}
\begin{theorem}
\label{theorem:stockmeyer}
For any classical probabilistic $T$-time
algorithm that outputs $z\in\{0,1\}^N$ with probability $q_z$,
and any constant integer $\alpha\ge1$,
there exists a
$O(T\log(T)\log\log(T))$-time classical probabilistic
algorithm that makes queries of length $O(T)$
to the NTIME[$n^2$] oracle
that 
outputs $\tilde{q}_z$ such that
\begin{eqnarray*}
|q_z-\tilde{q}_z|\le \frac{2^{\frac{1}{\alpha}}-2^{-\frac{1}{\alpha}}}{2}q_z
\end{eqnarray*}
with a success probability at least $w$
for any $z\in\{0,1\}^N$. Here, $w$ is a certain constant.
\end{theorem}

{\it Proof of Theorem~\ref{theorem:stockmeyer}.}
Let $C$ be a $T$-time deterministic classical algorithm
such that
\begin{eqnarray*}
\frac{|\{r\in\{0,1\}^T:C(r)=z\}|}{2^T}=q_z
\end{eqnarray*}
for all $z\in\{0,1\}^N$.
For each $z\in\{0,1\}^N$,
let us define the set $S_z\subseteq\{0,1\}^T$ by
\begin{eqnarray*}
S_z\equiv\{r\in\{0,1\}^T:C(r)=z\}.
\end{eqnarray*}
For any integer $\alpha\ge1$, define
\begin{eqnarray*}
S_z^{\times \alpha}\equiv
\{(r_1,...,r_\alpha)\in(\{0,1\}^T)^{\times \alpha}
:C(r_1)=...=C(r_\alpha)=z\}.
\end{eqnarray*}

For $S_z^{\times \alpha}$, find an integer $\eta\in\{1,2,...,\alpha T\}$
such that $A_{\eta-1}$ accepts and $A_\eta$ rejects
by the binary search.
It takes classical probabilistic $O(T\log(T)\log\log(T))$-time that makes
queries of length $O(T)$ to the
${\rm NTIME}[n^2]$ oracle.
(For the binary search, $O(\log T)$ steps are necessary, and each
implementation of the algorithm $A$ takes
$O(T\log\log(T))$ steps.)
Then, $2^{\eta-1}<|S_z^{\times \alpha}|<2^{\eta+1}$
with a sufficiently small failure probability.

If we define $\sigma\equiv 2^{\frac{\eta}{\alpha}}$,
$\frac{1}{2}\sigma^\alpha<|S_z|^\alpha 
<2\sigma^\alpha$.
Hence $\Big(\frac{1}{2}\Big)^{\frac{1}{\alpha}}\sigma<|S_z| 
<2^{\frac{1}{\alpha}}\sigma$.
If we define $\tilde{q}_z\equiv \sigma/2^T$,
we obtain the result.
\fbox

\end{document}